\documentstyle[12pt,epsfig]{article}

\begin{document}  
\vspace*{-2cm}  
\renewcommand{\thefootnote}{\fnsymbol{footnote}}  
\begin{flushright}  
hep-ph/9902428\\
DTP/99/22\\  
Febuary 1999\\  
\end{flushright}  
\vskip 65pt  
\begin{center}  
{\Large \bf Renormalization Group Effects in the Process
$H \longrightarrow \gamma \gamma$}\\
\vspace{1.2cm} 
{\bf  
Michael Melles\footnote{Michael.Melles@durham.ac.uk}  
}\\  
\vspace{10pt}  
Department of Physics, University of Durham,  
Durham DH1 3LE, U.K.\\  
  
\vspace{70pt}  
\begin{abstract}
The partial Higgs width $\Gamma (H \longrightarrow \gamma \gamma)$ is 
important at the LHC for Higgs masses in the MSSM mass window up to
$140$ GeV as a relatively background free signal of a fundamental scalar.
At the photon photon mode at the NLC it would be the Higgs production mechanism.
Two loop QCD corrections exist for the fermionic contribution and in the case
of the bottom loop large non-Sudakov double logarithms can be resummed to
all orders and contribute up to 12 \% compared
to the t-quark. In more complicated Higgs
sectors, such as in the MSSM, large $\tan \beta$ enhancements of bottom type
Yukawa couplings can potentially dominate even the whole partial width. A main uncertainty
in all existing calculations is the scale of the strong coupling as it is only
renormalized at the three loop level. In this paper we include the exact two
loop running coupling to all orders into the bottom contribution. We find
that the effective scale is close to $\alpha_s (10 m_b^2)$.
\end{abstract}
\end{center}  
\vskip12pt

\setcounter{footnote}{0}  
\renewcommand{\thefootnote}{\arabic{footnote}}  
  
\vfill  
\clearpage  
\setcounter{page}{1}  
\pagestyle{plain} 
 
\section{Introduction} 

The investigation of the electroweak symmetry breaking sector is certain to
dominate both theoretical as well as experimental high energy physics through
the coming decade. High precision measurements of electroweak observables at
the SLC and LEP indicate that the Standard Model Higgs boson has a mass
between $95$-$280$ GeV at the $95$\% confidence level with a preference
towards the lower mass regime \cite{f}. While the Standard Model (SM) has enjoyed spectacular
theoretical success, the Higgs sector of the theory is the only aspect for
which no ``standard'' in the experimentally tested sense has yet been
set. The simplest ansatz for the Higgs sector, assumed in the SM, leads to only 
one neutral fundamental scalar but several more complicated scenarios are
a priori just as viable. Well known examples include a general two doublet Higgs
model \cite{g,sh1,sh2} and the Higgs sector of the minimal supersymmetric extension
of the SM (MSSM) \cite{ghkd}. 

A common feature of all phenomenologically viable extensions of the SM Higgs
sector must be the existence of a ``SM-like'' neutral scalar in the 
SLC/LEP mass window. In the case of the MSSM, the lightest neutral scalar 
must be lighter than the Z-boson mass \cite{m}. Radiative corrections allow
its mass to reach at most $130$-$140$ GeV. The introduction of additional
singlet Higgs bosons can of course soften this upper bound somewhat, 
depending mainly on the value of the stop mass \cite{eh}, but at least
the MSSM would be ruled out for higher masses of the lightest neutral scalar.

Among the experimentally observable Higgs signals the $H \longrightarrow
\gamma \gamma$ coupling is of considerable interest. At the LHC it can be
used as a relatively background free process \cite{s1}. For Higgs masses
in the MSSM regime it is actually the only feasible signal due to the large QCD
background of other processes. Excellent energy and angular resolutions
of the detectors allow to discriminate the photon photon decay against the
continuum QCD background \cite{atlas,cms}. At the photon photon mode
of the NLC it would be the primary resonance production process
\cite{g1,g2,t}.

While the charged gauge boson loop gives the largest contribution in
the SM \cite{ghkd,s1}, the fermion loops deserve special attention. 
Radiation of a single gluon is not possible due to charge conjugation
invariance and color conservation. For the partial
width $\Gamma (H \longrightarrow \gamma \gamma)$ the fermion loops
interfere destructively and for large Higgs masses ($\sim$600 GeV) the top loop
nearly cancels the $W$ loop contribution. In that case the bottom loop would
be artificially ``enhanced'' in terms of its phenomenological importance.
There is of course also the possibility of $t$ and $b$ ``colluding'' in their
destructive interference with the charged gauge boson loop. 

In the SM, the top quark loop yields the bulk (almost 90\%) of the fermionic 
contribution to $\Gamma (H \longrightarrow \gamma \gamma)$ due to the large
Yukawa coupling. The bottom quark contribution is still significant 
(~12\%) due to
larger radiative corrections expounded on below \cite{s2}. 
In order to constrain new physics 
the theoretical SM-predictions should be as precise as possible.
One main source of uncertainty in the bottom quark contribution is the fact
that the strong coupling $\alpha_s$ receives no renormalization through two loops.
The same is of course also true for the top quark loop, however, the 
change in $\alpha_s$ between the two mass scales is much smaller and the
radiative corrections are also not as large in relative terms \cite{s2}. The 
width contribution stemming from a bottom quark, however, changes up to $17\%$ for
a Higgs boson mass below the $W^\pm$ threshold depending on whether one 
chooses to evaluate $\alpha_s$ at $m_b^2$ or $m_H^2$.

Tackling this uncertainty is the purpose of this work. There are two other
circumstances in which a renormalization group improvement is crucial. 
Firstly for the case 
of the second
heavy neutral scalar predicted in any model with two complex doublets.
In this scenario the aforementioned large radiative corrections would
be even bigger and the clear signal identification at a future collider
would hence benefit from a precise knowledge of QCD corrections.
The second and possibly much more important circumstance is that 
of a large $\tan \beta$ enhancement of bottom type Yukawa couplings.
These typically enter for two doublet Higgs sectors with a $\frac{1}{\cos
\beta}$ factor in the coupling \cite{ghkd}. Phenomenological restrictions
lead to a range roughly between $1 < \tan \beta < 50$ \cite{v}. As the width
is proportional to the square of $\frac{1}{\cos \beta}$, any enhancement
of the coupling could be substantial, even dominating.
Lastly it is also theoretically of interest to study the effect of 
renormalization group improvements for processes involving Yukawa couplings.

In the next section we briefly review the radiative corrections for the case
of the bottom contribution. Although an exact two loop calculation exists
\cite{sdgz}, the largest contribution is contained in non-Sudakov 
double logarithms (DL's)
and we will only review this calculation as it allows for an all orders 
inclusion of the renormalization group effects. The derivation of the running
coupling improved form factor is then given based on a topological similarity
between the $H \gamma \gamma$ and the Sudakov vertex in terms of their
double logarithmic content up to the last loop integration.
We then compare the renormalization group improved form factor with effective
couplings in the DL approximation and make concluding remarks about the
generality of the presented results.

\section{Renormalization Group Improved Form Factor} \label{sec:rg}

In this section we drop the index $m_b$ as it is unnecessary in that all
results derived are valid whenever the ratio $\frac{m^2}{m_H^2}$ leads to
large logarithms. For phenomenological applications we have the b-quark in
mind. We also explicitly only discuss the SM case. The only change for
more complicated Higgs sectors for the neutral scalars would be in the
Yukawa coupling. All 
DL-form factors and renormalization group effects would remain unchanged.

It was shown in Ref. \cite{ky} that in the case of a small ratio of $\frac{
m^2}{m_H^2}$ large non-Sudakov double logarithms occur in the amplitude
$H \longrightarrow \gamma \gamma$ via a fermion loop. The resulting series
can be expressed in the following form:
\begin{equation}
{\cal M}_q^{DL} (H \longrightarrow \gamma \gamma) = e^*_{1 \mu} e^*_{2 \nu}
\left(\frac{m^2_H}{2} g^{\mu \nu}-k^\mu_2 k^\nu_1 \right) \sqrt{ \sqrt{2} G_F}
\frac{e^2}{(4 \pi)^2} N_c \sum_q Q^2_q {\cal F}_H^{DL} \label{eq:adef}
\end{equation}
where $e_{1,2}$, $k_{1,2}$ are the polarization vectors and four-momenta of the
photons, $N_c$ the number of colors and $Q_q$ the electric charge of quark $q$
in units of the charge of the positron $e$.
The double logarithmic form factor is given by
\begin{equation}
{\cal F}_H^{DL} = \frac{4 m^2}{m_H^2} \log^2 \frac{m^2}{m^2_H} \sum^\infty_{n=0}
\frac{2 \Gamma (n+1)}{\Gamma (2 n +3)} \left( - \frac{\alpha_s C_F}{2 \pi}
\log^2 \frac{m^2}{m^2_H} \right)^n \label{eq:dlff}
\end{equation}
The leading order partial Higgs width was first given in Ref. \cite{egn}.
In order to improve the perturbative behavior of the quark loop contribution
one should use running quark masses with $m \left( \frac{m_H}{2} \right)$
as the effective scale for the photonic decay mode \cite{s1}.
Within the DL approximation the scale at which to evaluate the QCD-coupling
$\alpha_s$, however,
is unrestricted. Even for an exact calculation of the two $\gamma$ partial Higgs
width the strong coupling is only renormalized at the three loop level. 
As mentioned above, this inherent ambiguity leads to up to $17\%$ uncertainty
in the bottom contribution to the partial width $\Gamma (H \longrightarrow
\gamma \gamma)$ for an intermediate mass Higgs boson. 
In the following we will derive the
renormalization group effects of inserting a running coupling into each loop
evaluation for the exact one- and two-loop solution of the $\beta$-function. All
higher order RG-terms would then be suppressed by $\frac{1}{\log^3 \frac{s}
{m^2}}$. 

In the derivation of the leading logarithmic corrections
in Ref. \cite{ky} the familiar Sudakov technique \cite{Sud,gglf,fkm} was used
by decomposing loop momenta into components along external momenta 
, denoted by $\{\alpha, \beta\}$ and
those perpendicular to them, denoted by $l_\perp$. Integrating first over the
perpendicular components as usual,
the double logarithmic form factor of Eq. \ref{eq:dlff} was derived from a
double integral over the Sudakov form factor:
\begin{equation}
{\cal F}_H^{DL}= \frac{4 m^2}{m^2_H} \int^1_{-1} \int^1_{-1} \frac{d \alpha}{\alpha}
\frac{d \beta}{\beta} \Theta \left( \alpha \beta - \frac{m^2}{m^2_H} \right)
\exp \left[ - \frac{\alpha_s C_F}{2 \pi} \log |\alpha| \log |\beta| \right]
\label{eq:abdlff}
\end{equation}
\begin{center}
\begin{figure}
\centering
\epsfig{file=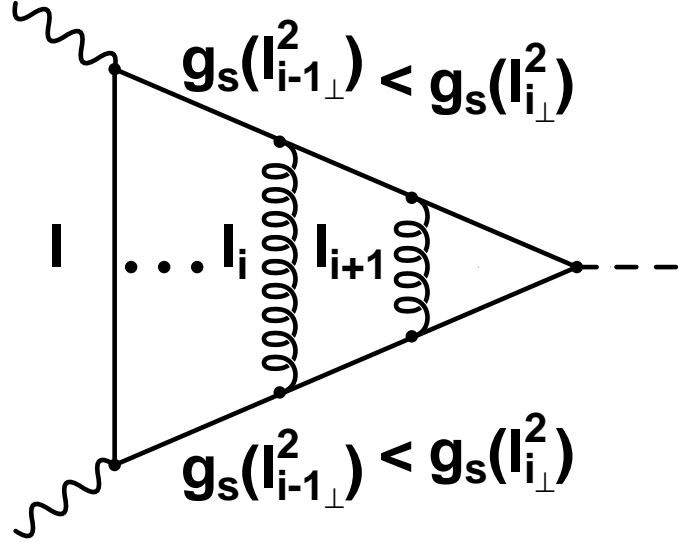,width=10cm}
\caption{The schematic Feynman diagrams leading to the hard (non-Sudakov)
double logarithms in the partial width $ \Gamma (H \longrightarrow 
\gamma \gamma )$ with $i+1$ gluon insertions. 
Crossed diagrams lead to a different
ordering of the Sudakov variables and are correctly accounted for by a factor
of $(i+1)!$ at each order. The scale of the coupling $\alpha_s=\frac{g_s^2}{4
\pi}$ is indicated at the vertices and included in this work.} 
\label{fig:HRG}
\end{figure}
\end{center}
We want to emphasize at this point that ideally one would like to have an
explicit three loop calculation in order to determine the effective scale
at the two loop level for a renormalization group improvement to all orders.
Without such a calculation is seems hard to attempt a rigorous inclusion of
the running coupling, however, the form of the derivation of the non-Sudakov
DL's in Eq. \ref{eq:abdlff} is the key point. 
For Sudakov double logarithms the relevant scale 
for the coupling at each loop
is $\alpha_s ( l^2_\perp )$ as was shown in
Refs. \cite{b,ddt,cl}. Following similar arguments we find that the same 
is true for
the novel non-Sudakov logarithms given in Eq. \ref{eq:dlff}.
One way to see this is
the fact that on a formal level, all insertions of gluons into the
DL-topology shown in Fig. \ref{fig:HRG} have the same structure as in the
Sudakov case. Eq. \ref{eq:abdlff} is the explicit mathematical formulation of
this fact. Only the last fermion loop integration separates the two cases
by effectively regularizing soft divergences with the quark mass. The strong
coupling receives no renormalization from this last integration,
however, so that
the scales of the couplings at each order are determined by the same
renormalization group arguments as for the Sudakov case.

\begin{center}
\begin{figure}
\centering
\epsfig{file=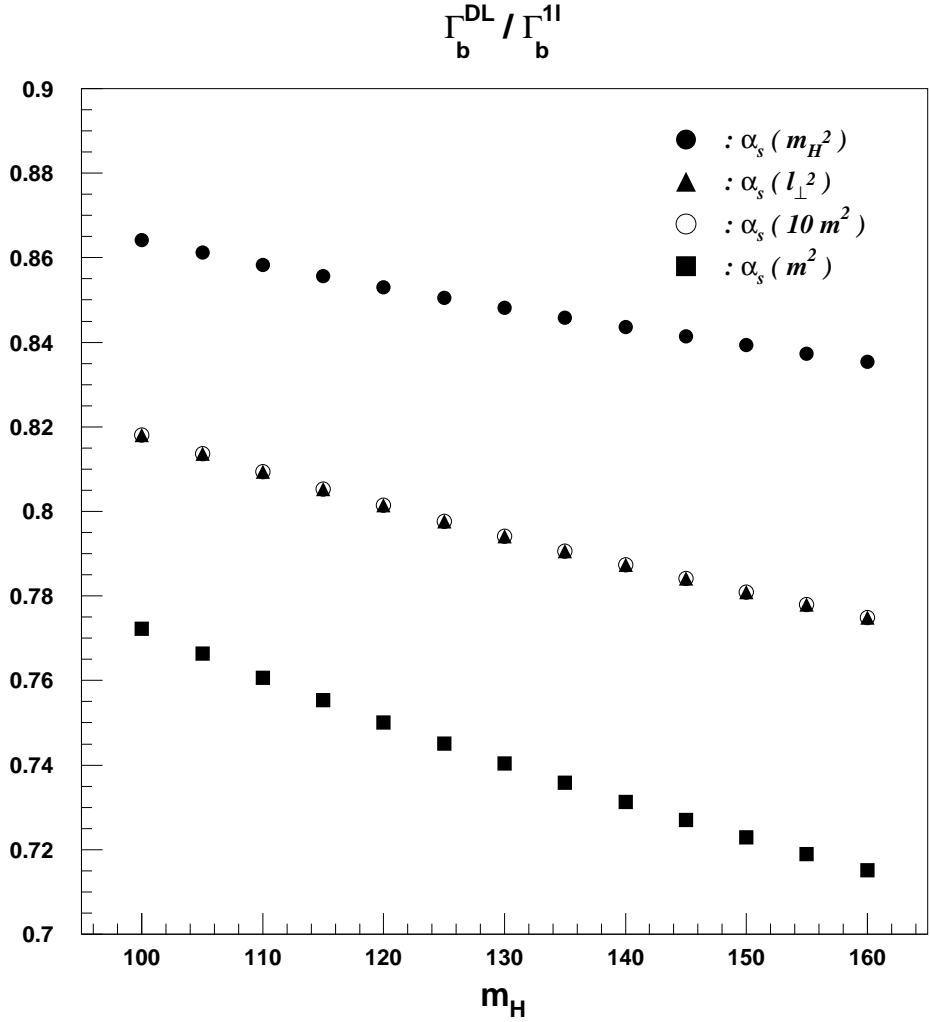,height=15cm}
\caption{The effect of incorporating a running coupling constant at each loop
integration (triangles) according to Eq. \ref{eq:rgw} 
to the partial Higgs width $\Gamma^{DL}_b ( H \longrightarrow \gamma
\gamma )$. Displayed is the ratio of the DL-bottom contribution to the width
relative to the one loop width.
Also shown are upper and lower
limits according to the indicated values of $\alpha_s$ in the DL approximation.
The renormalization group improved width is close to the effective scale
of using $\alpha_s (10 m^2_b)$ in the DL result (open circles). We use $\alpha_s(m^2_b)=0.2$
and $m_b=4.5$ GeV.}
\label{fig:hrg}
\end{figure}
\end{center}

We start by writing
\begin{equation}
\alpha_s (l^2_\perp) = \frac{\alpha_s(m^2)}{1+\beta_0 \frac{\alpha_s(m^2)
}{\pi} \log \frac{l^2_\perp}{m^2}+\beta_1 \left( \frac{\alpha_s(m^2)}
{\pi} \right)^2 \log \frac{l^2_\perp}{m^2}} \label{eq:rc}
\end{equation}
where $\beta_0=\frac{11}{12} C_A -\frac{4}{12} T_F n_F$, 
$\beta_1 = \frac{17}{24} C_A^2 - \frac{5}{12} C_A T_F n_F- \frac{1}{4} 
C_F T_F n_F$ and for QCD we have
$C_A=3$, $C_F=\frac{4}{3}$ and $T_F=\frac{1}{2}$ as usual. 
Up to two loops the massless $\beta$-function is independent of the 
chosen renormalization scheme 
and is gauge invariant in minimally subtracted schemes to all orders
\cite{c}. These features will also hold for the derived renormalization
group improved form factor below.
From the exact 
next to leading order result in Eq. \ref{eq:rc}
it is clear that a formulation in terms of $l^2_\perp$ of the series leading
to the non-Sudakov double logarithms is more adaptable to a renormalization
group improvement. Eq. \ref{eq:abdlff} can be reformulated  
in the following form:
\begin{eqnarray}
{\cal F}_H^{DL} \!\!\! 
&=& \!\!\!\! \frac{8 m^2}{m^2_H} \sum^\infty_{n=1}
\int^{m^2_H}_{m^2} \frac{d l^2_\perp}{l^2_\perp} \int^1_{\frac{l^2_\perp}{m^2_H}}
\frac{d \alpha}{\alpha} \prod^{n-1}_{i=1} \Gamma (n)  \int^{l^2_{{i-1}_\perp}}_{m^2} \frac{d l^2_{{i}_\perp}}
{l^2_{{i}_\perp}} \int^1_{\alpha_{i-1}} \frac{d \alpha_{i}}{\alpha_{i}} 
\left( \frac{-\alpha_s C_F}{2 \pi} \right)^{n-1} \label{eq:lpdl}
\end{eqnarray}
The product above is set to one for $n=1$ and contains nested integrals for 
$n \ge 2$ with $l^2_{{0}_\perp}
\equiv l^2_\perp$ and $\alpha_0 \equiv \alpha$.
From this expression it is clear that
an incorporation of the running coupling in Eq. \ref{eq:rc} will not contain 
any Landau pole singularity \cite{cm} as $m^2 \leq l^2_{i_\perp} \leq l^2_\perp$.
The fact that the strong coupling enters only at two loops means that for $i+1$
gluon insertions the first loop integral leads to a simple logarithm:
\begin{equation}
\int^{l^2_{i_\perp}}_{m^2} \frac{d l^2_{{i+1}_\perp}}{l^2_{{i+1}_\perp}}
= \log \frac{l^2_{i_\perp}}{m^2}
\end{equation}
This is a consequence of the requirement that the effective
scale be $l^2_{i_\perp}$
at the $i+1$-th gluon coupling. Fig. \ref{fig:HRG} indicates this schematically.
For the following integrals we use $c \equiv \left( \beta_0 \frac{ \alpha_s (m^2)
}{\pi} + \beta_1 \left( \frac{ \alpha_s (m^2)}{\pi} \right)^2 \right)$. At the
$i$-th order we then have:
\begin{equation}
\int^{l^2_{{i-1}_\perp}}_{m^2} \frac{d l^2_{{i}_\perp}}{l^2_{{i}_\perp}} \frac
{\alpha_s (m^2) \log \frac{ l^2_{{i}_\perp}}{m^2}}{1+c \;\log \frac{ l^2_{{i}_\perp}}{m^2}}
= - \frac{\alpha_s (m^2)}{c^2} \left( \log \frac{ \alpha_s (m^2)}{
\alpha_s (l^2_{{i-1}_\perp})} - c \log \frac{l^2_{{i-1}_\perp}}{m^2} \right)
\end{equation}
While the last term gives the same type of integral in the next step we use
for the first contribution on the r.h.s.: 
\begin{equation}
\frac{ d l^2_{i_\perp}}{l^2_{i_\perp}} =  
- \frac{\alpha_s(m^2)}{c} 
\frac{ d \log ( \alpha_s
(l^2_{i_\perp}))}{\alpha_s ( l^2_{i_\perp} )}
\end{equation}
and find
\begin{eqnarray}
&&-\int^{l^2_{{i-2}_\perp}}_{m^2} \frac{ d l^2_{{i-1}_\perp}}{l^2_{{i-1}_\perp}} \alpha_s (
l^2_{{i-1}_\perp}) \frac{\alpha_s(m^2)}{c^2} \left( 
\log \frac{\alpha(m^2)}
{\alpha(l^2_{{i-1}_\perp})} \right) \nonumber \\ &=& 
-\frac{\alpha^2_s(m^2)}{2 c^3} 
\log^2 \frac{\alpha(m^2)}{\alpha(l^2_{{i-2}_\perp})}
\end{eqnarray}
It is clear from this derivation at the n-loop level we have
\begin{equation}
\frac{1}{(n-1)!} \left(
\frac{\alpha_s(m^2) C_F}{2 \pi \; c} \right)^{n-1} 
\left( - \sum^{n-2}_{i=1} \frac{1}{i!}
\log^i \frac{\alpha_s (m^2)}{\alpha_s ( l^2_\perp )} + c \; \log \frac{ l^2_\perp 
}{m^2} \right) \frac{ \log^{n-1} \alpha}{\alpha}
\end{equation}
Thus we finally arrive at the complete renormalization group improved result
for the hard non-Sudakov form factor of
Fig. \ref{fig:HRG}:
\begin{eqnarray}
{\widetilde {\cal F}_H^{DL}} &=& \frac{4m^2}{m^2_H} \left[ \log^2 \frac{s}{m^2}-2 
\sum^{\infty}_{n=2} \int^s_{m^2} \frac{d l^2_\perp}{l^2_\perp } \left( \frac{
\alpha_s(m^2) C_F}{2 \pi \; c} \right)^{n-1} 
\frac{1}{n} \right. \nonumber \\
&&
\times \left. \left( - \sum^{n-2}_{i=1} \frac{1}{i!}
\log^i \frac{\alpha_s (m^2)}{\alpha_s ( l^2_\perp )} + c \; \log \frac{ l^2_\perp 
}{m^2} \right) \frac{\log^{n} \frac{l^2_\perp }{s}}{1+c \; \log \frac{
l^2_\perp }{m^2}} \right] \label{eq:rgdlff}
\end{eqnarray}
An expansion in $\alpha_s(m^2)$ gives the double logarithmic form factor in
Eq. \ref{eq:dlff} plus subleading terms proportional to $\beta_0$ etc..
The bottom contribution to the width is thus proportional to
\begin{equation}
\Gamma^{DL}_b (H \longrightarrow \gamma \gamma) \sim | {\widetilde {\cal F}_H^{DL}} 
| \; ^2 \label{eq:rgw}
\end{equation}
For precision predictions one should of course use the existing exact two loop
result from Ref. \cite{sdgz} up to that order, now, however, with the scale
uncertainty removed, and Eq. \ref{eq:rgw} for all 
higher contributions. 
The effect of the renormalization group improved form factor is shown in Fig.
\ref{fig:hrg} for the case of the b-quark. The effective scale of the 
coupling in the DL approximation close to $10 m_b^2$ and thus within a factor
of two to three lower than the mean logarithmic scale $m_H m_b$.
Fig. \ref{fig:hrg} also
displays that Eq. \ref{eq:rgw} will remain inside the upper and lower
DL-limits in the asymptotic regime.
We checked that by setting $\beta_1=0$ in $c$ we find almost the same
effective scale.
A similar effect can be found in 
the background process to $H \longrightarrow b \overline{b}$ at a photon
photon collider where also large non-Sudakov DL's enter \cite{jt1,jt2,fkm,ms1,ms2}. In this
case the effect of the renormalization group improvement is larger 
as it already occurs at the two loop level but the effective scale in that
process is of the same order of magnitude \cite{ms3}. 

\section{Conclusions} \label{sec:con}

The partial Higgs width $\Gamma ( H \longrightarrow \gamma \gamma )$ is essential
for the experimental discovery of a neutral scalar Higgs boson 
below 140 GeV and its production at
the photon photon mode of the NLC. Large QCD non-Sudakov double logarithms enhance
the bottom contribution relative to the top in the Standard Model up to 12 \%
but it may be much larger if more complex Higgs sector scenarios are realized
in nature. 
The main uncertainty for existing
QCD corrections remained in the scale of the coupling, leaving up to a
17 \% uncertainty in the bottom contribution. In this paper we have included
the exact running coupling through two loops and find that the effective
scale is given roughly by $\alpha_s (10 m^2_b)$, in between the lower 
and upper 
theoretically allowed regime. The renormalization group improved form factor
in Eq. \ref{eq:rgdlff} is gauge and scheme independent. 
The analysis is also valid for heavier Higgs masses as well as
the neutral Higgs bosons of general two Higgs doublet models including the
MSSM. Phenomenological implications of this
result are clearly more significant for the latter as the bottom loop can be
substantially enhanced through large Yukawa couplings.

\noindent{\bf Acknowledgements}\\ 
We would like to thank W.~Beenakker for discussions.
This work was supported in part by the EU Fourth Framework Programme `Training and Mobility of 
Researchers', Network `Quantum Chromodynamics and the Deep Structure of Elementary Particles', 
contract FMRX-CT98-0194 (DG 12 - MIHT).

\end{document}